\documentclass[9pt,preprint2,natbib209]{aastex6}

\begin{document}


\title{Effects of the core-collapse supernova ejecta impact on a rapidly rotating massive companion star}
\author{ Chunhua Zhu\altaffilmark{1}, Guoliang L\"{u}\altaffilmark{1},
 Zhaojun Wang\altaffilmark{1}}
\email{$^\dagger$zhuchunhua@xju.edu.cn, guolianglv@xao.ac.cn}
\altaffiltext{1}{School of Physical Science and Technology, Xinjiang
University, Urumqi, 830046, China.}





\begin{abstract}
We investigate the effects of the core-collapse supernova ejecta on a rapidly rotating and massive companion star.
We show that the stripped mass raises by twice when compare with a massive but non-rotating companion star.
In close binaries with orbital periods of about 1 day, the stripped
masses reach up to $\sim 1 M_\odot$. By simulating the evolutions of the rotational velocities of
the massive companion stars based on different stripped masses, we find that the rotational velocity
decreases greatly for stripped mass that is higher than about $1 M_\odot$.
Of all the known high mass X-ray binaries (HMXBs), Cygnus X-3 and 1WGA J0648.024418
have the shortest orbital periods of 0.2 and 1.55 days, respectively.
The optical counterpart of the former is a Wolf-Rayet star,
whereas it is a hot subdwarf for the latter.
Applying our model to the two HMXBs,
we suggest that the hydrogen-rich envelopes of their optical
counterparts may have been stripped by CCSN ejecta.
\end{abstract}

\section{Introduction}
High mass X-ray binaries (HMXBs) are composed of a compact object (CO) and a massive companion star of
OB spectral type whose matter is accreted onto the CO. The CO is usually a neutron star (NS) or a black hole (BH).
HMXBs are divided into BeHMXBs
and sgHMXBs. In the former, the companions of accreting COs are Be stars, whereas they are supergiants in the latter.
Of the 114 HMXBs known in the Galaxy \citep{Liu2006}, about 32\% are sgHMXBs and more than 60\% are BeHMXBs. Supergiant fast X-
ray transients (SFXTs) are a subclass of sgHMXBs. They are characterized by sporadic, short and intense X-ray fares. There are about 10 SFXTs known in the Galaxy \citep[e. g.,][]{Romano2014}. In the Small Magellenic Cloud, 148 are confirmed candidates of HMXBs \citep{Haberl2016}, in which only SMC X-1 belongs to sgHMXBs while others are BeHMXBs. \cite{Antoniou2016} classified 40 HMXBs in the Large Magellenic Cloud and found 33 BeHMXBs and 4 sgHMXBs, including two systems in which the COs may be BHs. Usually, the progenitors of HMXBs undergo core-collapse supernovae (CCSNe). When COs form, CCSN ejecta collides with their companion stars. It was shown by many literatures that the interaction of CCSN ejecta with the companion star may affect the evolution of the latter if the orbital periods are short enough \citep[e. g.,][]{Wheeler1975,Hirai2014,Liu2015}. Using two-dimensional (2D) hydrodynamical simulations, \cite{Hirai2014} found that up to 25\% of the companion¡¯s mass can be stripped for the shortest binary separations, whereas \cite{Liu2015}  suggested that at most 10\% of the companion¡¯s mass is removed based on 3D hydrodynamical simulations. The stripped mass depends greatly on binary separations.

\begin{figure}
\includegraphics[totalheight=3.0in,width=3.5in,angle=-90]{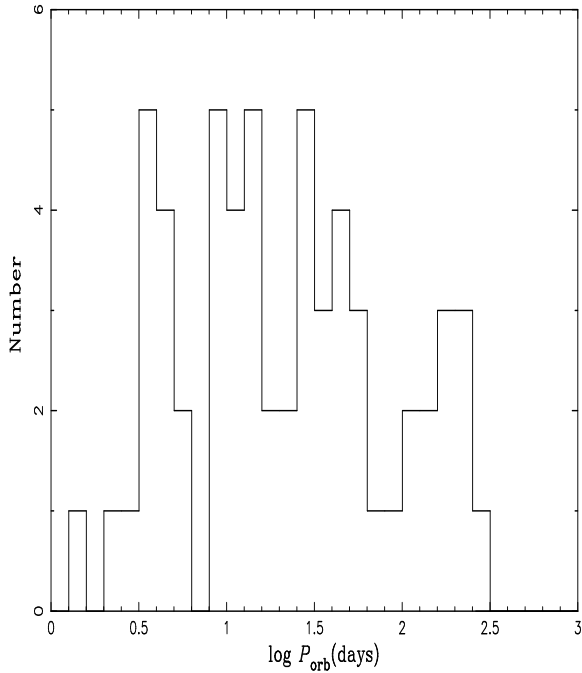}
\caption{ Orbit periods of HMXBs in the Milk Way based on the observational data from \citet{Liu2006} and \citet{Romano2014}. }
\label{fig:porb}
\end{figure}

Figure \ref{fig:porb} shows the distribution of the Galactic HMXBs¡¯ orbital periods. Obviously, there are 17 HMXBs with orbital period shorter than 10 days. In particular, the orbital period of Cygnus X-3 is only 0.2 days \citep{Parsignault1972}, and its optical counterpart, V1521 Cyg, is a Wolf-Rayet star \citep{Kerkwijk1992}. Similarly, 1WGA J0648.024418 has an orbital period of 1.55 days \citep{Thackeray1970} with an optical counterpart of a hydrogen-depleted subdwarf O6 (HD 49798). In the remaining 15 HMXBs, their optical counterparts are supergiants, that is, they are sgHMXBs \citep{Liu2006, Romano2014}. Based on \cite{Wheeler1975}, \cite{Hirai2014} and \cite{Liu2015}, these optical counterparts should be greatly affected by the ejecta of supernovae. However, to our knowledge, the effects of CCSN ejecta on the companion star are seldom considered in previous literatures on HMXBs. As \cite{Shao2014} showed, the companion stars may have been spun up and become Be stars before COs were formed in HMXBs with short orbital periods. When CCSNe occur, a certain amount of mass should be removed from these Be stars. We ask the questions: How does the surface velocity of these Be stars evolve? Are these Be stars remain Be stars after the redistributions of angular momentum?

In this paper, we investigate the impact of CCSN ejecta on their rotating massive companion
stars, and try to understand the optical counterparts of HMXBs. \S 2 describes the stellar model we use for simulating the evolution of rotating stars. The stripped masses from rotating massive stars during CCSN are estimated in \S3, and the evolutions of rotating massive stars after stripped masses are shown in \S 4. We present the main conclusions in \S5.

\section{Be Stars and The Evolution of Rotating Stars}
Be stars are B type stars with very high rotational velocity, close to the critical velocity ($V_{\rm crit}$)
where gravity is balanced by the centrifugal force.
A review of Be stars can be seen in \cite{Porter2003}. In HMXBs, Be stars, which originate from
main sequence (MS) stars, accrete matter from their companion stars. A detailed investigations and their properties of Be stars
can be found in the works by \cite{Mink2013}, \cite{Shao2014} and \cite{Reig2011}.
The range of mass distribution for Be stars
in HMXBs is between 8 and 22 $M_\odot$ \citep{Chaty2013}. For simplicity, in this work,
we define phenomenologically a MS star as Be star if its mass is between  8 and 22 $M_\odot$ and
its rotational velocity is higher than 80\% of $V_{\rm crit}$ \citep{Porter2003}.

Rotation has significant effects on the massive stars \citep{Maeder2000}.
Here, we use MESA (version 8118, \citealt{Paxton2011,Paxton2013,Paxton2015}) to compute the
structures and evolutions of rotating massive stars. By considering
the physics of rotation, mass loss and magnetic fields, \cite{Brott2011} gave
the grids of evolutionary models for rotating massive stars. In order to compare with
their results, we use similar input parameters and criteria in MESA:
The Ledoux criterion
is used for convection, mixing-length parameter ($\alpha_{\rm LMT}$) is taken as 1.5,
an efficiency parameter ($\alpha_{\rm SEM}$) of unity is assumed  for
semi-convection. The metallicity ($Z$) of the Milk Way is taken as 0.0088 in \cite{Brott2011}, which
is lower than that of the Sun ( $Z_\odot = 0.012$ ) found by \cite{Asplund2005}.

The mass-loss rate is calculated via the model
given by \cite{Vink2001}. Due to rotation, the mass-loss rate is enhanced and given by\citep{Langer1998}
 \begin{equation}
\dot{M}=(\frac{1}{1-\Omega/\Omega_{\rm crit}})^\beta \dot{M}^{0},
\label{eq:ml}
\end{equation}
where $\Omega$ and $\Omega_{\rm crit}$ are the angular velocity and the critical angular velocity, respectively,
and $\beta=0.43$\citep{Langer1998}.

Rotation induces instability of various kinds, such as dynamical shear instability, Solberg-Hi{\o}land instability,
secular shear instability, Eddington-Sweet circulation, and the
Goldreich-Schubert-Fricke instability\citep[e.g.,][]{Heger2000}, which results in
the transport of angular momentum \cite[e.g.,][]{Endal1978,Pinsonneault1989,Heger2000}. Following
\cite{Brott2011}, the ratio of the turbulent viscosity to the diffusion coefficient ($f_{\rm c}$) is taken as 0.0228 \citep{Heger2000},
and the ratio of sensitivity to chemical gradients ($f_{\rm \mu}$) as 0.1 \citep{Yoon2006}.

\begin{figure}
\includegraphics[totalheight=3.0in,width=3.5in,angle=-90]{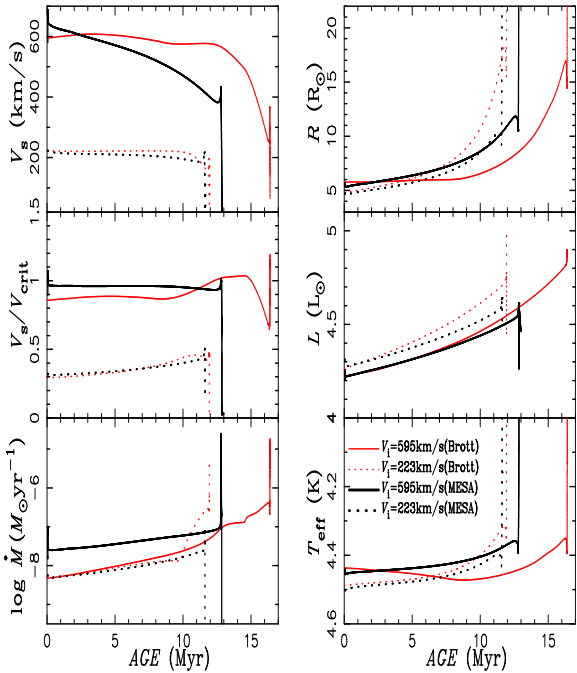}
\caption{The evolutions of physical parameters for a star with initial mass of 15
$M_\odot$ on the MS phase. $V_{\rm s}$ and $V_{\rm crit}$ are the rotational velocity on the stellar surface
and the critical rotational velocity, respectively. $V_{\rm i}$ represents the initial value of
$V_{\rm s}$. Different $V_{\rm i}$ is represented by different
linetypes which are shown in the legend in the bottom right plot.
The brackets indicate that the data are either taken from the grids of \cite{Brott2011} or calculated by MESA.
 }
\label{fig:mebr}
\end{figure}

Figure \ref{fig:mebr} shows the evolution of a star with initial mass of 15 $M_\odot$ on the MS phase.
It is obvious that, for the models with low initial rotational velocity ($V_{\rm i}=223$ km s$^{-1}$), the results calculated by MESA and \cite{Brott2011}
are in excellent agreement although there are small differences in luminosity and radius.
However, this is not the case in models for high $V_{\rm i}$ of 595 km s$^{-1}$, where the results of MESA are
quite different from those in \cite{Brott2011}. Especially, when $V_{\rm i}$ increases
from 223 to 595 km s$^{-1}$, \cite{Brott2011} predicted a prolong in the lifetime of MS by about
30\% whereas a prolong of only about 10\% is predicted in MESA. Furthermore,  the rotational velocity ($V_{\rm s}$) calculated by MESA
decreases more rapidly than that using the model of \cite{Brott2011}.   In this work, we do not discuss the details that 
result in these differences, but one should note that they may lead to some large uncertainties in simulating
rapidly rotating massive stars.

\section{Core-Collapse Supernovae}

\subsection{Effects of Core-Collapse Supernovae on Orbital Periods}
Due to an asymmetry during CCSN, a new-born NS receives a kick velocity ($v_{\rm k}$) which may disrupt the binary system.
\cite{Brandt1995} systematically studied the effects of the kick velocity on the orbital periods.
They found that post-CCSN distributions for binary parameters, such as orbital period, eccentricity,
are determined by numerous factors including  pre-CCSN orbital period, eccentricity and stellar masses,
post-CCSN stellar masses, the magnitude and direction of the kick velocity.
As shown in Figure 1 in \cite{Belczynski2008}, the pre-CCSN masses of NS progenitors are between
about 6 and 10 $M_\odot$ using the mass-loss rates in \cite{Nieuwenhuijzen1990},
\cite{Kudritzki1978} and \cite{Vassiliadis1993} for H-rich stars on MS, red giant and asymptotic giant branches,
 respectively. For simplicity,
we assume the pre-CCSN masses to be 8 $M_\odot$ for all NS progenitors, and
1.4 $M_\odot$ as the mass for all new-born NSs \citep{Belczynski2008,Lattimer2007}.
 As binary systems usually have undergone binary interaction, such as mass transfer or tidal interaction,
   the pre-CCSN eccentricity before CCSNe occur is taken as 0.
Of course, a new-born BH also obtains a kick velocity. However, there is no observational evidence for it.
Therefore, we assume that the kick velocities for BHs are similar to those of NSs. Based on Figure 1 in \cite{Belczynski2008},
we take 10 $M_\odot$ as the pre-CCSN masses for all BH progenitors, and the masses for all new-born BHs as
8 $M_\odot$.
Hence, in this paper, the binary parameters for post-CCSN systems are assumed dependent on the magnitude and direction of the kick velocity.

Based on the measured proper motions for 233 pulsars, \cite{Hobbs2005} found that
the distribution of kick velocities can be perfectly described by a Maxwellian distribution
\begin{equation}
P(v_{\rm k})=\sqrt{\frac{2}{\pi}}\frac{v^2_{\rm k}}{\sigma^3_{\rm
k}}e^{-v^2_{\rm k}/2\sigma^2_{\rm k}}.
\label{eq:kick}
\end{equation}
with a dispersion of $\sigma_{\rm k}=265$km s$^{-1}$.
This implies that the direction of kick velocity is uniform over all solid angles.

For a given pre-CCSN binary,  we can determine
the distribution for values and directions of kick velocities using Monte Carlo method,
and calculate the probability of survival for a system after CCSN, and its binary parameters.
A detailed description and the codes for the above calculations can be found in \cite{Hurley2002}.

\begin{figure}
\includegraphics[totalheight=3.0in,width=3.5in,angle=-90]{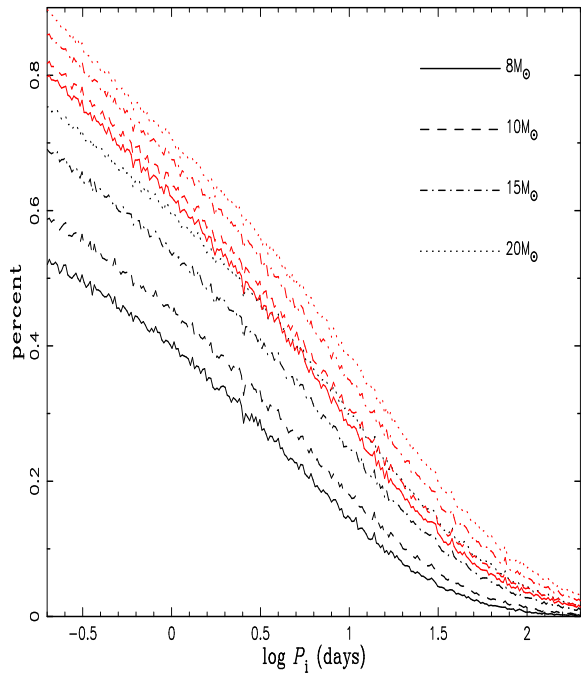}
\caption{The percentage of bound remaining binaries after CCSNe. $P_{\rm i}$ is
the pre-CCSN orbital period. The black and red lines represent that COs are the NSs and
BHs, respectively.}
\label{fig:undr}
\end{figure}

In this work, we use the codes provided by \cite{Hurley2002} to
calculate the binary parameters of post-CCSN systems.
Considering that the orbital periods of HMXBs are between 0.2 and 300 day (See Figure \ref{fig:porb}),
we take a similar range for the orbital periods ($P_{\rm i}$) of pre-CCSN systems.
 Figure \ref{fig:undr} shows the percentage of bound remaining in binaries after CCSNe.
  Here, $\Delta \log P_{\rm i}=0.1$ day, and 10000 binary systems are calculated for every orbital period.
 The results show that the larger the orbital period for a pre-CCSN binary is, the more easily the binary is
disrupted. The smaller the masses for the companions of CO' progenitors in pre-CCSN binaries are, the more difficult for the binaries
to survive. Since the masses of new-born BHs ($\sim 8 M_\odot$) are larger than those of new-born NSs ($\sim 1.4 M_\odot$),
hence, for the same orbital periods, the binaries that produce BHs will remain bounded more readily than those that produce NSs.

\begin{figure*}
\begin{tabular}{lr}
\includegraphics[totalheight=3.0in,width=3.5in,angle=-90]{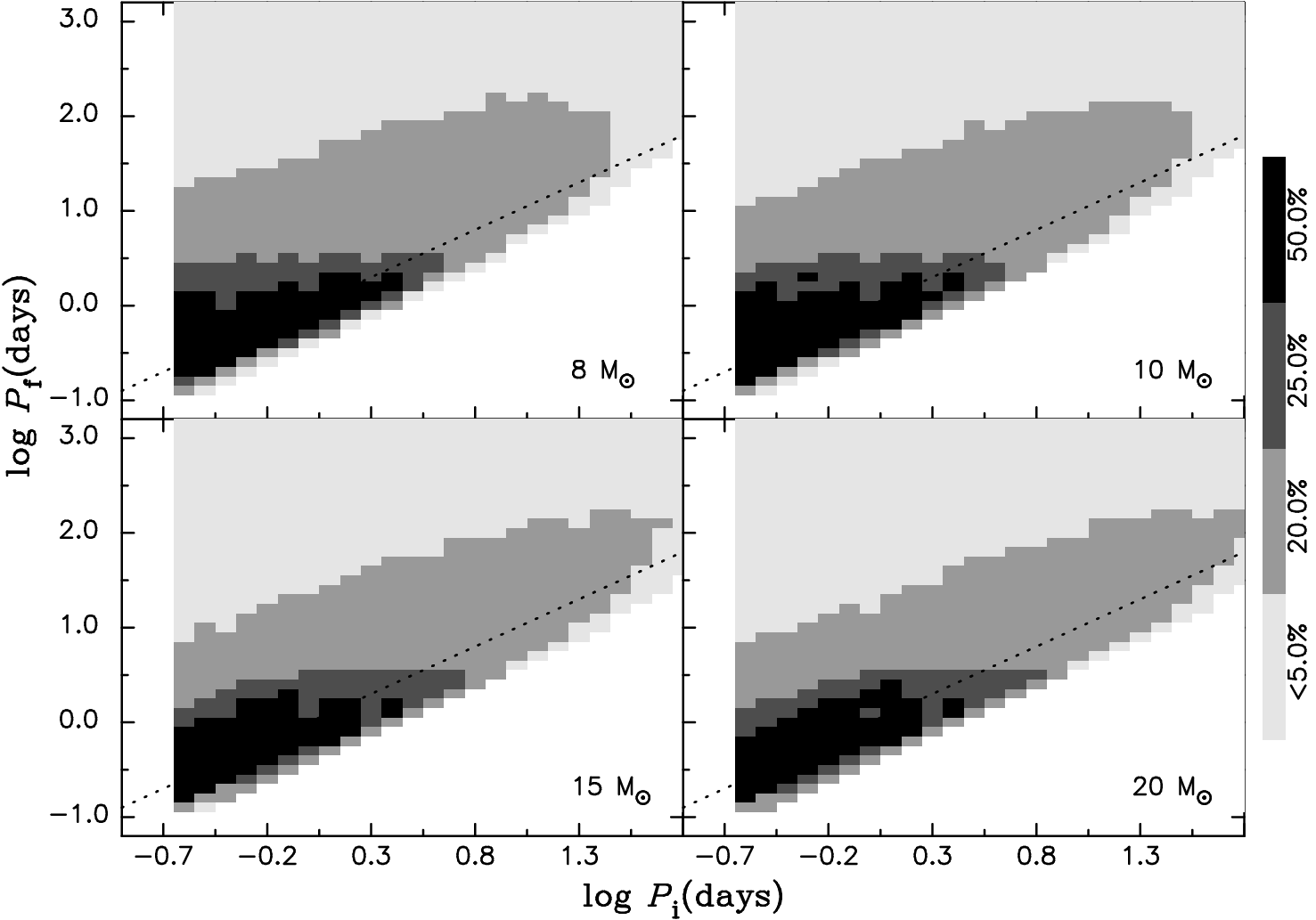}&
\includegraphics[totalheight=3.0in,width=3.5in,angle=-90]{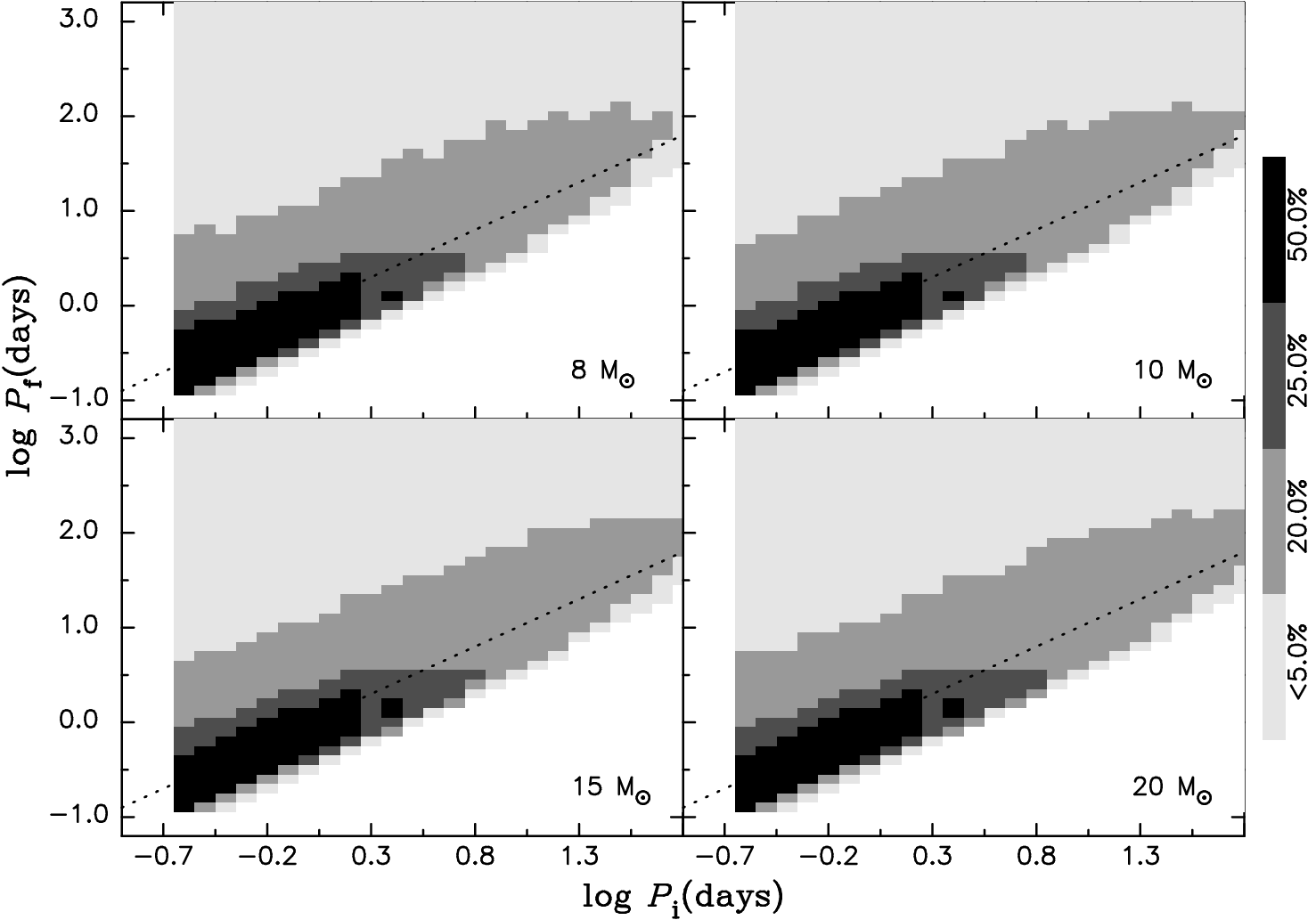}\\
\end{tabular}
\caption{The distribution of pre- and post-CCSN orbital periods.
The COs are NSs and BHs in the left and right panels, respectively.
 $P_{\rm i}$ is the pre-CCSN orbital period, while
$P_{\rm f}$ is the post-CCSN orbital period. The dotted line indicates
$P_{\rm i}=P_{\rm f}$.
}
\label{fig:pfpi}
\end{figure*}

Figure \ref{fig:pfpi} gives the distribution of pre- and post-CCSN orbital periods.
It is clear that the range of orbital period in a post-CCSN binary is wider.
The post-CCSN binaries with orbital periods shorter than 10 days may originate
from pre-CCSN binaries with orbital periods shorter than about 30 days.
By comparing the left panel to the right panel in Figure \ref{fig:pfpi},
the changes in orbital periods from pre- to post-CCSN are similar regardless of whether CCSN produces
 BHs or NSs.

\subsection{Impact of Core-collapse Supernovae on Rotating Massive Stars}
CCSNe do not only affect the orbital periods of binaries, but also have impact
on the companion stars of new-born CO progenitors
\citep[e.g.,][]{Wheeler1975,Marietta2000,Pan2012,Shappee2013,Liu2015}.
At the beginning of CCSN ejecta colliding with the companion stars, a shock is sent into the stellar envelope
of the latter, and, simultaneously, a reverse shock is sent back into the former.
The shock propagates throughout the companion star, while the reverse shock turns into a bow shock around
the companion star. The results of the impact of the CCSN ejecta on the companion star is
that much of the shock energy is deposited in the companion's envelope which turns into internal energy causing the
companion star to heat up. If the internal energy
of the material in the companion's envelope is high enough, it is stripped away from the companion star.

The stripped mass is determined by local total energy given by \cite{Marietta2000},
\begin{equation}
E_{\rm tot}=E_{\rm kin} + E_{\rm in} + E_{\rm gr},
\label{eq:et}
\end{equation}
where $E_{\rm kin}$, $E_{\rm in}$ and $E_{\rm gr}$ are the specific kinetic
energy, the specific internal energy, and the specific potential energy, respectively.
The first and second are positive, while the third is negative. Matter is stripped if $E_{\rm tot}>0$.

\cite{Pan2012} found that the final stripped mass can be estimated using the power law given by
\begin{equation}
M_{\rm st}=A\left(\frac{a}{R_2}\right)^\eta M_2,
\label{eq:mst0}
\end{equation}
where $R_2$ and $M_2$ are radius and mass of the companion star, respectively, and $a$ is the binary separation.
Here, $A$ and $\eta$ are fitting parameters whose values are dependent on the properties of CCSNe
(including energy, mass and velocity of ejecta), the structure of companion
stars (including radius, density profile) and binary separations \citep[e.g.,][]{Pan2012}.
 Both \cite{Hirai2014} and \cite{Liu2015} estimated the stripped masses from the companion stars during CCSNe.
The former focused on the red-giant companion stars, while the latter investigated the MS companion stars.
The detailed structures of a red-giant companion star are very different from stars in the MS phase.
Therefore,  $A=0.26$ and $\eta=-4.3$ are taken in \cite{Hirai2014}, while they are $0.143$ and $-2.65$, respectively, in \cite{Liu2015}.
Although we focus on MS stars, we take the value for fitting parameters ($A$ and $\eta$) from both in \cite{Hirai2014} and \cite{Liu2015}
to estimate the stripped masses in order to discuss the effects of $M_{\rm st}$ on rotating massive stars.

Simultaneously, the shock heating can change the internal structures of the companion stars.
However, it is beyond the scope of
this work to calculate $M_{\rm st}$ by 2D or 3D hydrodynamical simulations and to simulate
the change of stellar structures due to the shock heating. MESA cannot simulate stripping process
but can be used to calculate the evolution of a star with high
mass-loss rate.
Following \cite{Podsiadlowski2003}, we
assume that the impact of CCSNe on the companion star can be
divided into two phases.
In the first phase, the companion star loses mass at a very high rate($\sim 10^{-2}-10^{-3} M_\odot$yr$^{-1}$)
until the mass lost equals $M_{\rm st}$ given by Eq. (\ref{eq:mst0}).
This means that its thermodynamic equilibrium is destroyed at such high mass-loss rate. The result is that both
the stellar radius and the rotational velocity at the stellar surface decrease as the stellar mass decreases.
In the second phase,
the mass loss stops but the companion star is irradiated by external heating source until $E_{\rm tot}=0$ at the
stellar surface. In fact, similar work was done by \cite{Shappee2013} using MESA code but without considering the stellar rotation.
 Here, we must note that it is still different even when an additional heating source
 is introduced to simulate the shock heating due to the interaction between SN ejecta and a companion star.
As shown by \cite{Pan2012}, \cite{Liu2013} and \cite{Hirai2014},
the internal structures of a star are strongly affected while the shock is passing through the star.
However, in our model, the internal structures of massive stars are not affected by such interaction.

\begin{figure}
\includegraphics[totalheight=3.0in,width=3.5in,angle=-90]{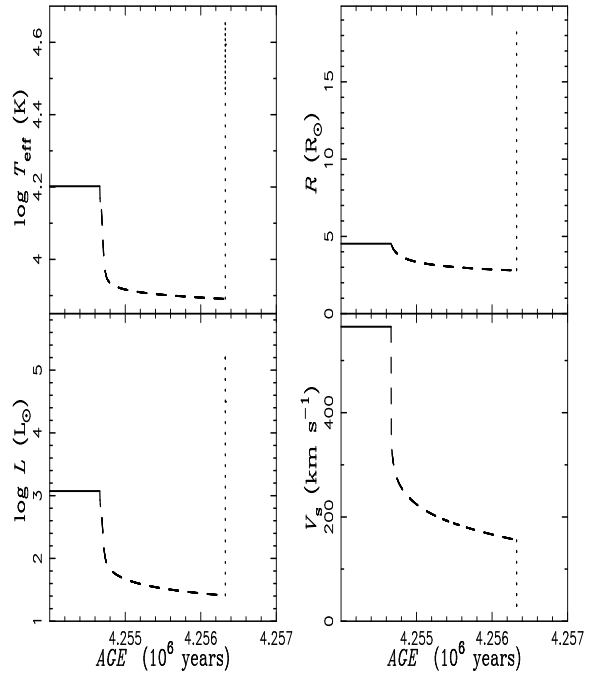}
\caption{The evolution of the MS companion star with mass of 10 $M_\odot$ during different phases.
The solid, dashed and dotted lines represent the normal phase, mass-loss phase with a
very high rate and irradiated phase, respectively. See text for more details.
}
\label{fig:itev}
\end{figure}

Figure \ref{fig:itev} shows an example for the evolutions of a MS companion star with mass of 10 $M_\odot$
in the different phases mentioned above.
When the central hydrogen abundance of the MS decreases to 90\% of its initial value (its age is
about $4.2547\times10^6$ year), the impact of CCSN begins. Based on our assumptions, the MS loses
mass at a rate of $\sim$$10^{-3}M_\odot$ yr$^{-1}$ at first. If the mass-loss rate is higher,
MESA code stops due to the convergence problem. During this phase, the effective temperature,
the stellar radius, the stellar luminosity and the rotational velocity drop along with the mass lost.
When the stellar mass reduces to 9$M_\odot$, the mass-loss phase stops and the MS enters
the irradiated phase. In this example, a value of $10^{20}$erg s$^{-1}$ cm$^{-2}$ is assumed for the energy flux
that irradiates the MS from the heating source, which means that the power of total irradiation energy is about
$4.3\times10^{43}$erg s$^{-1}$ at the beginning of the irradiated phase. As Figure \ref{fig:itev} shows,
the stellar radius increases rapidly as a result of the irradiation, which leads to great enhancement on the power.
With the increase of the radius and the temperature, $E_{\rm in}$ becomes higher and higher but $|E_{\rm gr}|$
becomes smaller and smaller around the stellar surface.
After about 8000 seconds (equivalent to energy of about $10^{46}$ erg deposited into the stellar envelope ),
the $E_{\rm tot}\geq 0$ at the stellar surface and the irradiated phase stops.


According to \cite{Hirai2014} and \cite{Liu2015}, the stripped mass from the companion star
of CCSN progenitor is determined by its mass, radius, rotating velocity and binary separation.
As shown in Figure \ref{fig:mebr}, both the radius and rotating velocity
change with stellar evolution. Therefore, the stripped masses depend indirectly on the stellar age.
We use the mass fraction of central hydrogen ($XH$) to represent the stellar age, where
$XH=0.9XH_{\rm i}$ and $XH=0.1XH_{\rm i}$ represents young and old stars, respectively.
Using MESA, we simulate evolution for stars with initial masses of 8, 10, 15 and 20$M_\odot$, respectively.
Their initial rotational velocities on the stellar surface are 0.95 $V_{\rm crit}$, 0.85 $V_{\rm crit}$ and 0.
We select two points in time, at $XH=0.9XH_{\rm i}$ and $XH=0.1XH_{\rm i}$,
on MS phase for discussion of the effects of stellar evolutions. Table \ref{tab:radii}
gives the radii of the MS stars for different masses and rotation velocities at the two time points.

\begin{table*}
\centering
\begin{minipage}{176mm}
  \caption{The radii of the MS stars for different masses and rotation velocities
  at $XH=0.9XH_{\rm i}$ and $XH=0.1XH_{\rm i}$, respectively. The first column gives the stellar mass when CCSNe occur.
  The stellar ages are given in columns 2 and 6. The stellar radii are shown from columns 3 to 5 and 7 to 9.
  The numbers in square brackets from columns 3 to 5 and 7 to 9 are the rotational velocity.
  }
\tabcolsep2.0mm
\begin{tabular*}{160mm}{|c|c|c|c|c|c|c|c|c|}
\cline{1-9}
\multicolumn{1}{|c|}{}&\multicolumn{4}{|c|}{$XH=0.9XH_{\rm i}$}&\multicolumn{4}{|c|}{$XH=0.1XH_{\rm i}$} \\
\cline{2-9}
\multicolumn{1}{|c|}{$M_{\rm }$}&\multicolumn{1}{|c|}{$AGE$}&\multicolumn{3}{|c|}{$R (R_\odot)\ \ [V_{\rm s}\ ({\rm km \ s^{-1}})]$}&\multicolumn{1}{|c|}{$AGE$}&\multicolumn{3}{|c|}{$R (R_\odot)\ \ [V_{\rm s}\ ({\rm km \ s^{-1}})]$}\\
\cline{3-5}\cline{7-9}
\multicolumn{1}{|c|}{($M_\odot$)}&\multicolumn{1}{|c|}{($10^6$yr)}& $V_{\rm s}=0$ &$V_{\rm s}=0.85V_{\rm crit}$&$V_{\rm s}=0.95V_{\rm crit}$&\multicolumn{1}{|c|}{($10^7$yr)}& $V_{\rm s}=0$ &$V_{\rm s}=0.85V_{\rm crit}$&$V_{\rm s}=0.95V_{\rm crit}$\\
 \cline{1-9}
 8&6.0&3.44\ [0]&4.06\ [491]&4.15\ [543]&2.9&6.22\ [0]&7.56\ [351]&9.79\ [387]\\
 10&4.3&3.95\ [0]&4.63\ [507]&4.74\ [560]&2.0&7.24\ [0]&8.78\ [355]&9.04\ [390]\\
 15&2.1&4.97\ [0]&5.80\ [538]&5.93\ [594]&1.1&9.69\ [0]&11.73\ [358]&12.09\ [391]\\
 20&1.5&5.86\ [0]&6.81\ [554]&6.97\ [613]&0.8&12.11\ [0]&14.83\ [346]&15.30\ [380]\\
  \cline{1-9}
\end{tabular*}
\label{tab:radii}
\end{minipage}
\end{table*}

\begin{figure*}
\begin{tabular}{lr}
\includegraphics[totalheight=3.0in,width=3.5in,angle=-90]{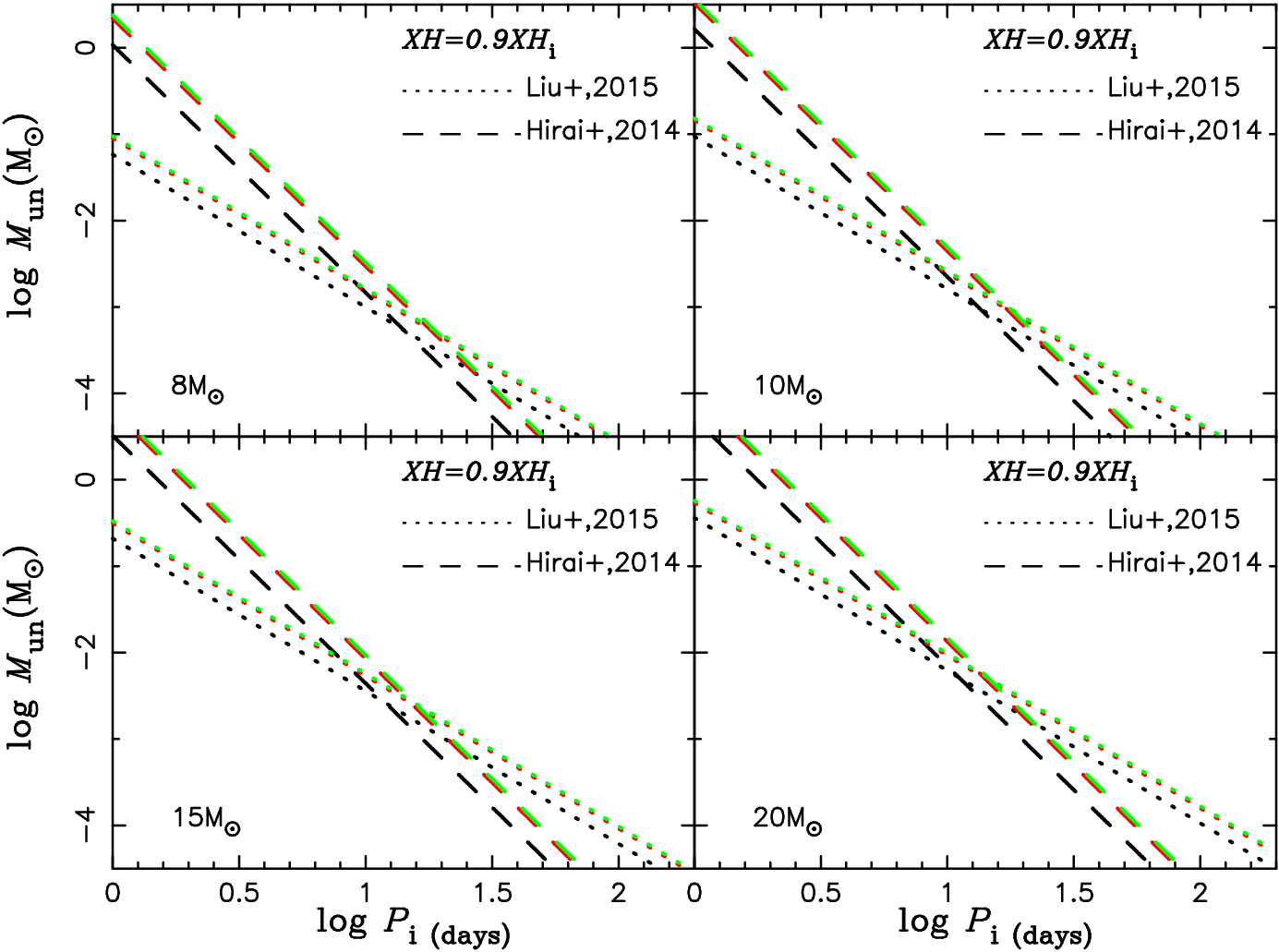}&
\includegraphics[totalheight=3.0in,width=3.5in,angle=-90]{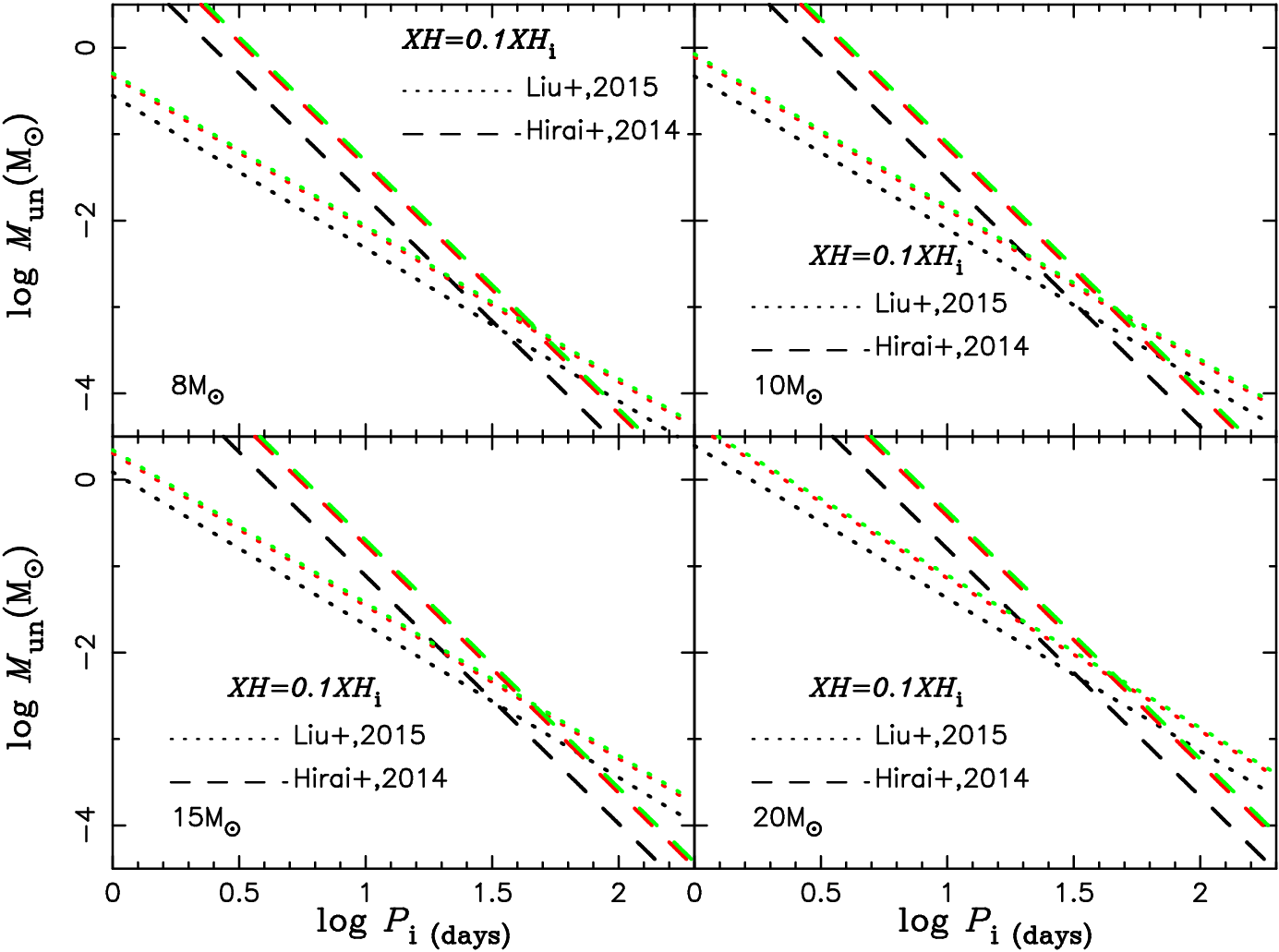}\\
\end{tabular}
\caption{Stripped masses ($M_{\rm un}$) vs. pre-CCSN orbital periods. The dashed and dotted lines
represent the fitting parameters $A$ and $\eta$ taken from
\citet{Hirai2014} and \citet{Liu2015}, respectively. Lines of the same type but in different colors signify $V_{\rm i}=0$ (Black),
$0.85V_{\rm crit}$ (red) and $0.95V_{\rm crit}$ (green), respectively.
The left and right panels represent different occurring time for CCSNe when the mass fractions of central hydrogen of companion stars
are 0.9 and 0.1$XH_{\rm i}$, respectively, where $XH_{\rm i}$ is the initial mass fractions of central hydrogen.
}
\label{fig:umass}
\end{figure*}

Assuming that the mass of NS progenitor in CCSN is 8$M_\odot$ allows us to estimate the stripped masses
from these stars in binary systems with different pre-CCSN orbital periods.
Based on Figure \ref{fig:umass}, compared to the model for non-rotating stars, the stripped mass in the model for
rapid rotation stars ($V_{\rm s}=0.95V_{\rm crit}$) is enhanced within a factor of about 2.3.
As shown in Table \ref{tab:radii}, the stellar radii of the former
is about 1.3 times larger than that of the latter, which results in the increase of 1.6 and 2.3 for
the fitting parameters in \cite{Liu2015} and \cite{Hirai2014}, respectively (See Eq. (\ref{eq:et})).
Compared to young massive stars ($XH=0.9XH_{\rm i}$) with high rotational velocity, CCSN ejecta can strip more matter from
the evolved massive stars ($XH=0.1XH_{\rm i}$) because the stellar radius increases by a factor of about 2 from the time point of
$XH=0.9XH_{\rm i}$ to that of $XH=0.1XH_{\rm i}$. According to
our calculations, the stripped masses in the model with $XH=0.1XH_{\rm i}$ are about 6---20 times larger than that with $XH=0.9XH_{\rm i}$.

The stripped mass greatly depends on the orbital period. Figure \ref{fig:pfpi} shows that HMXBs with orbital periods shorter than
10 days originate from pre-CCSN binaries with orbital periods shorter than 30 days.
Based on the Figure \ref{fig:umass},
we find that, regardless of whether using fitting parameters in \cite{Hirai2014} or using those in \cite{Liu2015},
the stripped masses from rotating massive stars in pre-CCSN binaries with orbital period shorter than 30 days
are larger than $\sim 10^{-3} M_\odot$, and even up to several $M_\odot$.

\label{sec:umass}
\section{Evolution of Rotating Massive Stars after the Impact}

According to \cite{Hirai2014} and \cite{Liu2015}, the timescale, $t_{\rm st}$, for mass stripping during CCSN,  is
several hours. The distribution of angular momentum within the star may depend on the Eddintton-Sweet
circulation \citep{Zahn1992}, whose timescale is given by
\begin{equation}
t_{\rm ES}\sim t_{\rm KH}(\frac{\Omega_{\rm crit}}{\Omega})^2,
\end{equation}
where $t_{\rm KH}$ is the local thermal timescale. It is apparent that
$t_{\rm ES}\sim t_{\rm KH}$ for Be stars ($\Omega \sim \Omega_{\rm crit}$).
For massive stars on MS phase, $t_{\rm KH}\sim 10^4-10^5$ years \citep{Heger2000}, which means that
$t_{\rm ES}$ is much longer than $t_{\rm st}$.
Therefore, the internal profiles of rotational velocity and angular momentum for rotating
stars do not change when their matter is stripped during CCSNe.
After a certain amount of mass is stripped from a star, its thermodynamic equilibrium is disrupted.
A new equilibrium will be reached after an adjustment within a thermal timescale, during which
the angular momentum within the star redistributes. After the rotating star reaches a new thermodynamic equilibrium, it begins to
evolve as a non-Be star.

\begin{figure*}
\begin{tabular}{lr}
\includegraphics[totalheight=3.0in,width=3.5in,angle=-90]{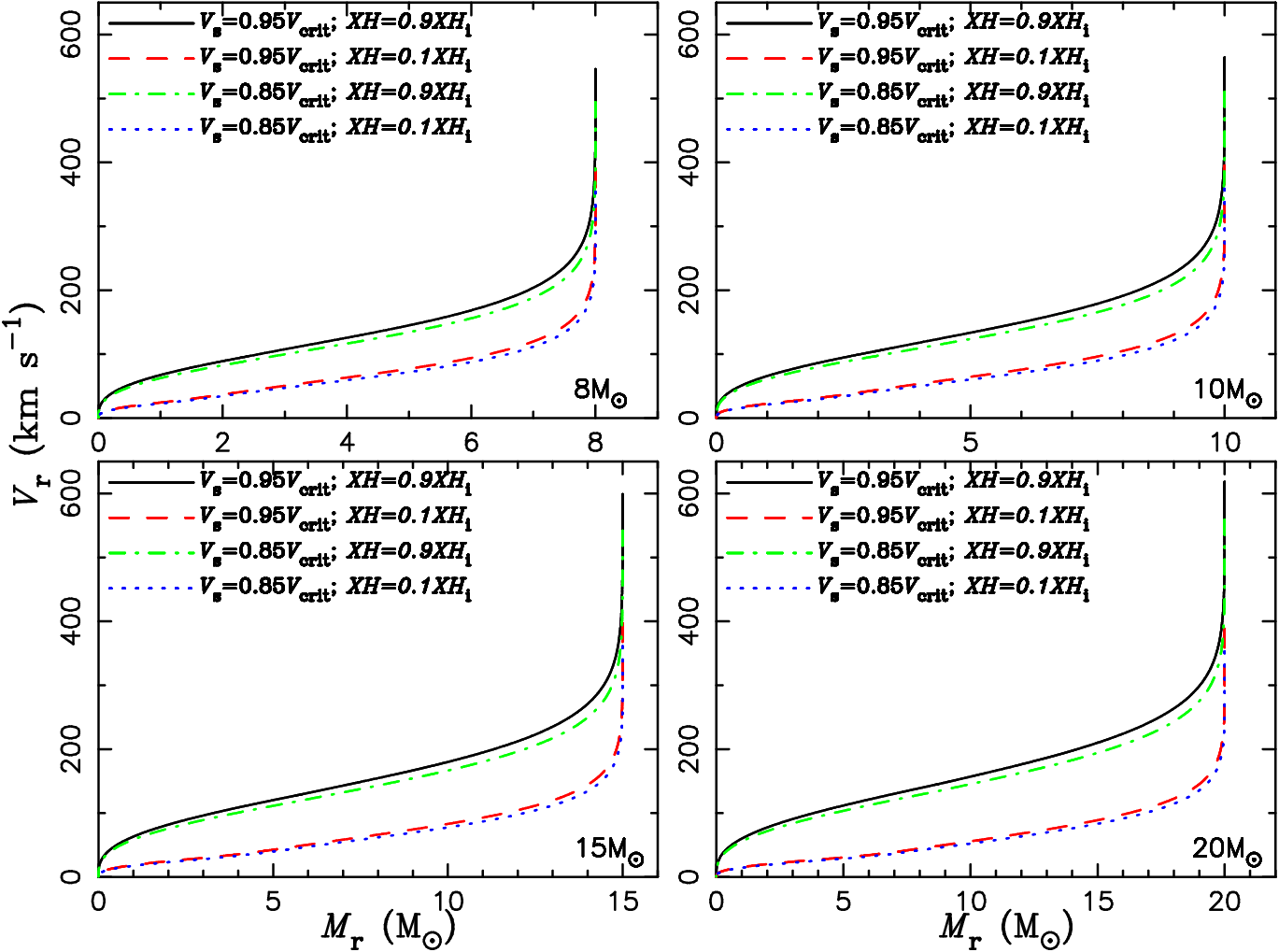}&
\includegraphics[totalheight=3.0in,width=3.5in,angle=-90]{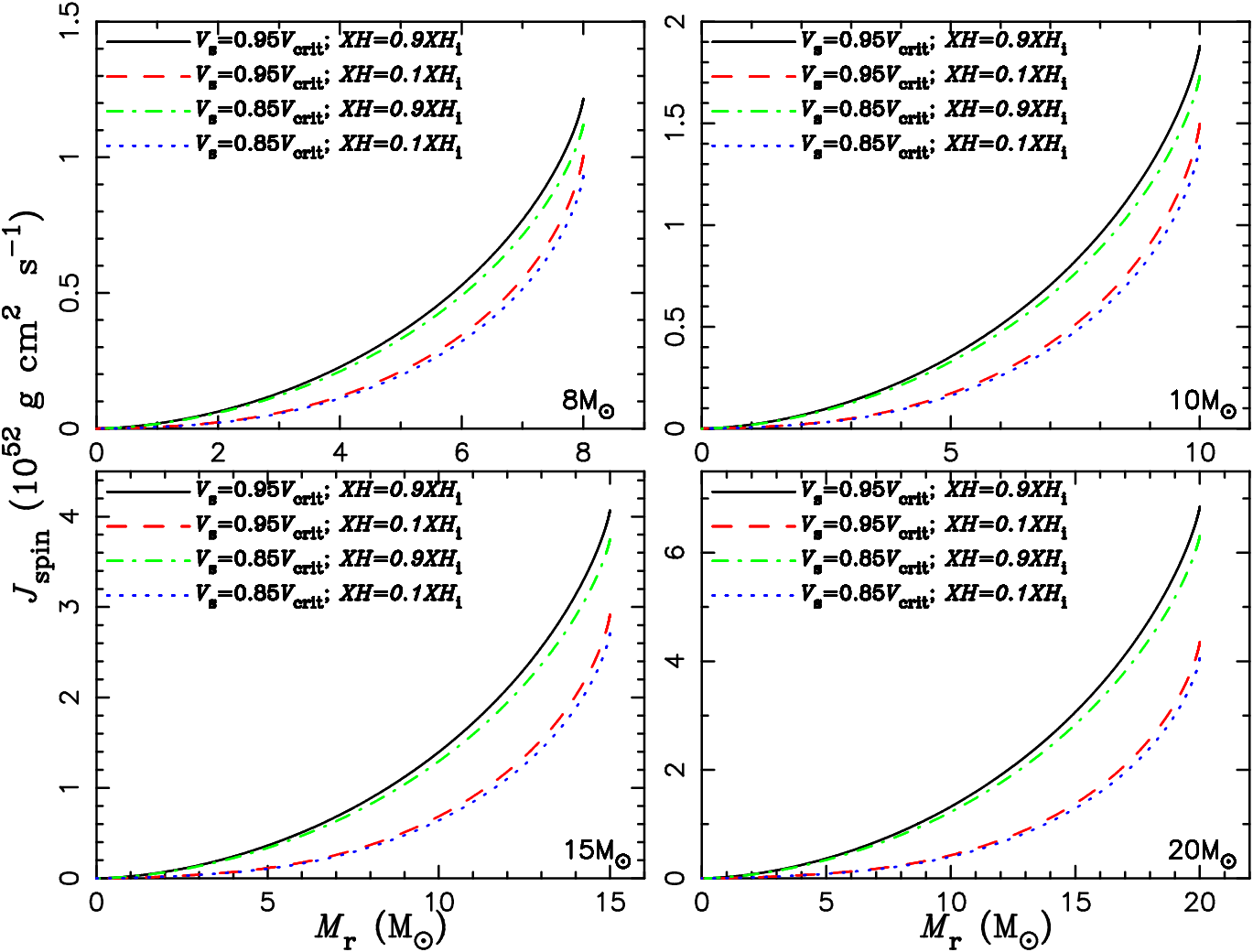}\\
\end{tabular}
\caption{The internal profiles for rotational velocity (left panel) and angular momentum (right panel)
vs. mass, that coordinates from stellar surface to center, of different stellar mass types with different rotational velocities
and evolutionary ages. }
\label{fig:jspin}
\end{figure*}

\subsection{Evolution of Rotational Velocity}
Figure \ref{fig:jspin} shows the internal profiles of rotational velocity and angular momentum ($J_{\rm spin}$) from
stellar surface to center. It is obvious from the figure that the rotational velocity decreases rapidly 
when the stellar-mass coordinates from the surface to the sub-surface. This means that the rotational velocity on the stellar surface
also decreases as the stellar matter being stripped. From the left panels in Figure \ref{fig:jspin},
a rapidly rotating star ($V_{\rm s}=0.95V_{\rm crit}\sim600$km s$^{-1}$) turns into a non-Be star with low rotational
velocity ($V_{\rm s}\sim 300$km s$^{-1}$) even though only a mass of $10^{-3} M_\odot$ is stripped away.
However, stellar rotational velocity depends on the stellar angular momentum.
Compared the right panel with the left panel in Figure \ref{fig:jspin}, the degree of reduce in
angular momentum is much lower than that in the rotational velocity if a certain mass is stripped away from a star.
A problem appears: Is a Be star still a Be star after a certain amount of mass is stripped?

In order to answer this problem, we investigate the rotational velocity evolution of rotating star after a certain mass is stripped.
In this work, we roughly divide the rotational velocity evolution of rotating star stripped mass
into three phases:\\
(i)Impact phase. This phase includes the stripped and the irradiated phases described in \S 3.2  \\
(ii)Thermally adjusting phase. After the impact, the heating source disappears, and the star undergoes adjustment to
reach a new thermodynamic equilibrium. This phase
lasts for a thermal timescale. According to \cite{Heger2000}, the secular shear instability, Eddington-Sweet circulation and the
Goldreich-Schubert-Fricke instability begin to drive the distribution of angular momentum on a thermal
timescale, and they are secular processes. Therefore, during this phase, the above three instabilities do not work, but
 dynamical shear instability and Solberg-Hi{\o}land instability affect the distribution of angular momentum.  \\
(iii)Normal phase. The rotating star begins to evolve into a non-Be star after it reaches a new thermodynamic equilibrium.
   \\

\begin{figure}
\includegraphics[totalheight=3.0in,width=3.5in,angle=-90]{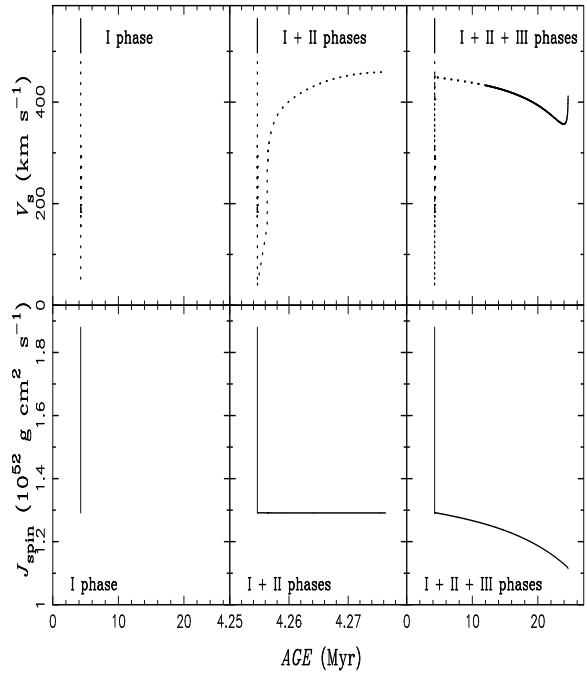}
\caption{The evolution of rotational velocity on the stellar surface and angular momentum of star during
impact, thermally adjusting and normal phases for a star of $10 M_\odot$  with stripped mass of $1 M_\odot$.
The left panel is for phase I (the impact phase),
the middle panel is for phases I and II (the impact and the thermally adjusting phases),
and the right panel is for phases I, II and III ( the impact, the thermally adjusting and the normal phases).
The solid lines on the top three panels represent Be star ( $V_{\rm s}>0.8V_{\rm crit} $ ), while
the dotted lines represent non-Be star ($V_{\rm s}<0.8V_{\rm crit} $).
}
\label{fig:111}
\end{figure}

Figure \ref{fig:111} gives the evolution of $V_{\rm s}$ and $J_{\rm spin}$ for
the above three phases for a  star of $10 M_\odot$ with stripped mass of $1 M_\odot$.
At the beginning, the star is on MS phase with the abundance of hydrogen at the center ($XH$) is
0.9 times of the initial hydrogen abundance ($XH_{\rm i}$), and $V_{\rm s}=0.95 V_{\rm crit}$.
That is, the star is a Be star. As given in Table \ref{tab:radii}, the age, radius and rotational
velocity of the star are, respectively, about $4.3\times10^6$ yr, 4.74 $R_\odot$ and 560 km s$^{-1}$ at this time,
and, based on Figure \ref{fig:umass}, it may exist in a binary system with the orbital period shorter than 1 day.

\begin{figure}
\includegraphics[totalheight=3.0in,width=3.5in,angle=-90]{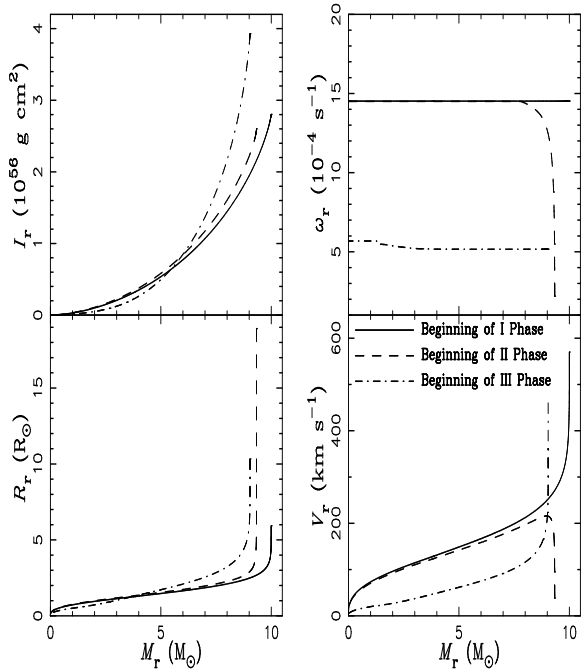}
\caption{The internal profiles for  physical parameters
vs. mass coordinate ($M_{\rm r}$) from stellar surface to center at the beginning of the three phases
for a star of $10 M_\odot$ mass and stripped mass of $1 M_\odot$. The parameters $I_{\rm r}$,
$\omega_{\rm r}$, $R_{\rm r}$ and $V_{\rm r}$ are the moment of inertia, angular velocity, radius and rotational velocity at
$M_{\rm r}$, respectively. The phases I, II and III indicate the impact, the thermally adjusting and the normal phases, respectively.
}
\label{fig:stru}
\end{figure}

At the beginning of the impact phase,
the Be star has a high mass-loss rate (about $10^{-2}-10^{-3}M_\odot$yr$^{-1}$)
so that the stellar angular momentum cannot be redistributed in a short
timescale of about $10^{2}-10^{3}$ yr. Therefore, $V_{\rm s}$ and $J_{\rm spin}$ decrease rapidly as the stripped matter increases.
Meanwhile, the stellar
radius also reduces quickly enhancing the critical rotational velocity. Soon, the star
is no longer a Be star but turns into a non-Be star. At the end of the stripped phase, the $V_{\rm s }$  and
$J_{\rm spin}$ decrease from about 560 to 200 km s$^{-1}$ and from about $1.87\times10^{52}$ to $1.29\times10^{52}$ g cm$^2$ s$^{-1}$,
respectively. Our simulation shows that the high mass-loss phase only lasts for hundreds or thousands of years.
After that, the star, having been stripped with a mass of 1 $M_\odot$, is irradiated by a heating source. Its envelope
rapidly expands, and its radius sharply increases while the rotational velocity on the surface drops radically.
However, as Figure \ref{fig:stru} shows, the irradiation only affects the structure near the stellar surface.

Entering the thermally adjusting phase, the mass-loss rate reduces down to normal value (about $10^{-8}M_\odot$yr$^{-1}$) and the irradiation stops.
The star begins to contract reaching a new thermodynamic equilibrium. From  Figure \ref{fig:stru}, its radius reduces from about $20 R_\odot$ to
10 $R_\odot$.
Due to the low mass-loss rate and the relative short timescale, $J_{\rm spin}$ remains almost a constant implying that
the angular velocity decreases by 2.3 times.  However, the star remains in the solid-body rotation
because of the existence of Spruit-Tayler magnetic fields, which results in the increase of $V_{\rm s}$.
This phase lasts for about $10^4$ years.

After the rapidly expanding phase, the star begins to evolve into a non-Be star. The $J_{\rm spin}$
and $V_{\rm s}$ decrease because of the matter lost taking away the angular momentum. However, as the star expands,
$V_{\rm crit}$ decreases more quickly than $V_{\rm s}$. Then, at about $1.2\times10^7$ years,  when $V_{\rm s}>0.8V_{\rm crit}$,
the star becomes a Be star.


\begin{figure*}
\begin{tabular}{lr}
\includegraphics[totalheight=3.0in,width=3.5in,angle=-90]{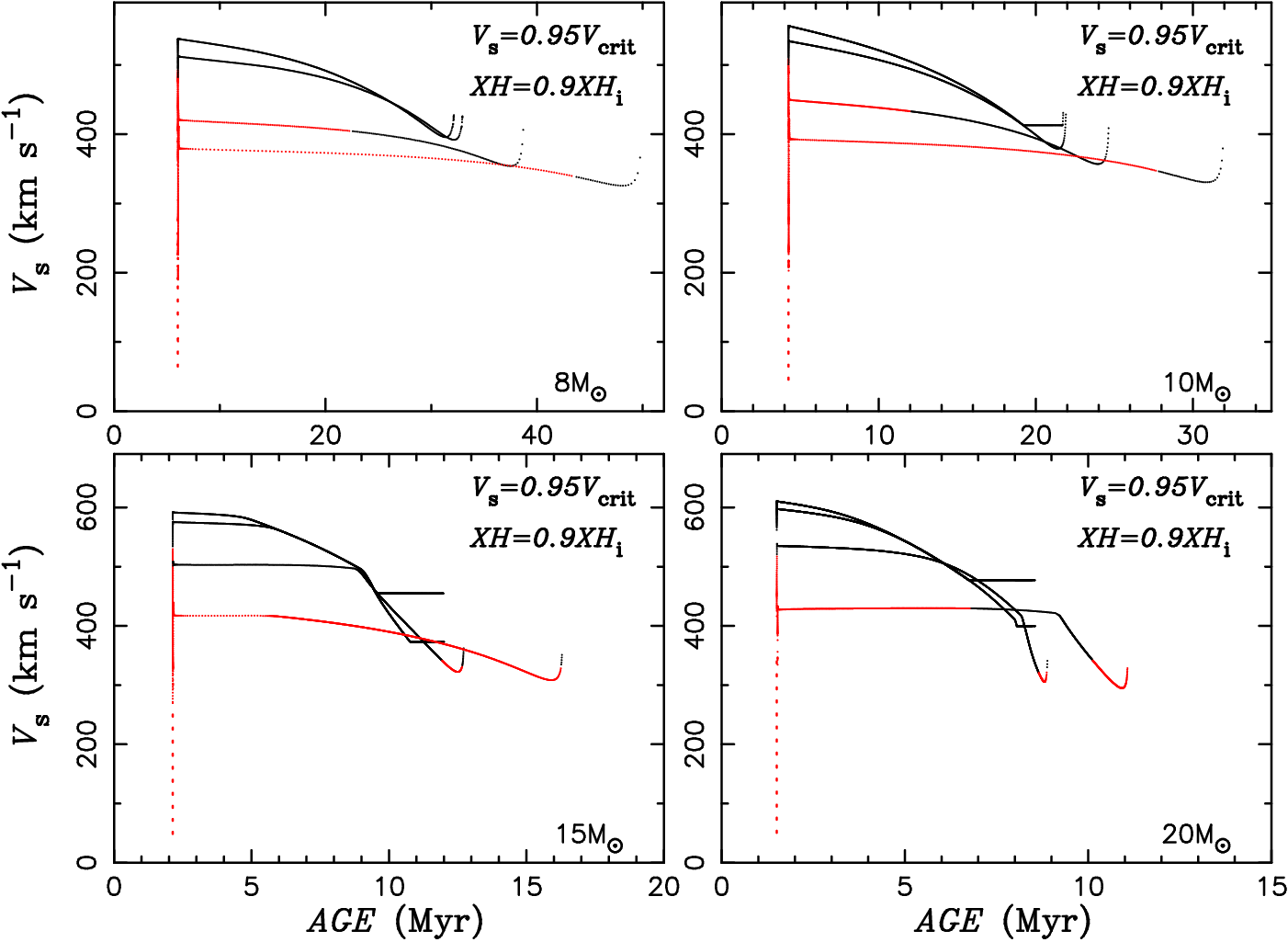}&
\includegraphics[totalheight=3.0in,width=3.5in,angle=-90]{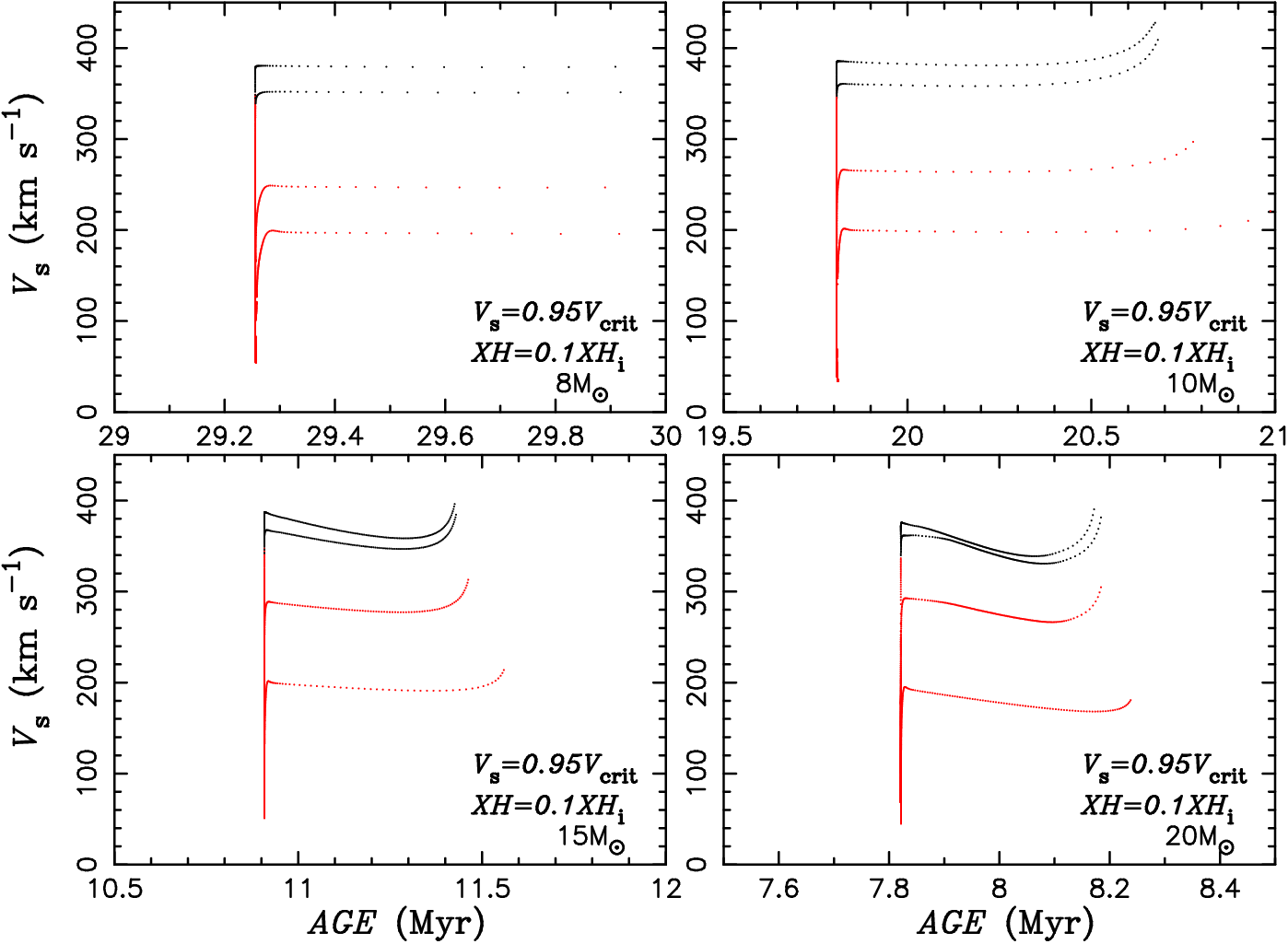}\\
\end{tabular}
\caption{The evolutions of rotational velocity on the stellar surface for stars with initial
masses of 8, 10, 15 and 20 $M_\odot$. The left and right panels are for
different evolutionary ages at $XH=0.9XH_{\rm i}$ and $XH=0.1XH_{\rm i}$,
respectively. The four dotted lines represent, from the top to
the bottom , $10^{-3} M_\odot$, $10^{-2} M_\odot$, $1 M_\odot$ and 25\% of stellar mass stripped, respectively.
The black and red dots mean that the stars are Be stars ($V_{\rm s}>0.8V_{\rm crit}$) and non-Be stars
($V_{\rm s}<0.8V_{\rm crit}$), respectively. Every dot represents a model calculated by MESA. }
\label{fig:avte}
\end{figure*}

We calculate the evolution of $V_{\rm s}$ for stars with stripped masses
of $10^{-3} M_\odot$, $10^{-2} M_\odot$, $1 M_\odot$ and 25\% of stellar mass at different
evolutionary ages. Compare to the old Be stars given in the right panel of Figure \ref{fig:avte},
the young Be stars in the left panel of Figure \ref{fig:avte} are more difficult to turn into non-Be stars
for the same stripped mass.
The main reason is shown in the right panel of Figure \ref{fig:jspin}:
the ratio of the angular momentum taken away by the matter stripped near the stellar surface
to the total angular momentum of the old Be stars is higher than that for the young Be stars.
The same reason can also be used to explain why Be stars with higher mass are more difficult to evolve into non-Be stars
than their lower mass counterpart. In short, a Be star with a certain amount of mass stripped can hardly evolve into a non-Be star unless
the stripped mass is larger than 1 $M_\odot$, even 25\% of its mass.

\subsection{Discussions }
As shown in Figure \ref{fig:umass}, a significant amount of mass
($\sim$ several $M_\odot$) should have been stripped from the progenitors of CO
companions during CCSNe for HMXBs with very short orbital periods ($\sim 1$ days).
Therefore, these companions may be hydrogen-depleted objects.
In known HMXBs, the orbital period of Cygnus X-3 is the shortest ($P_{\rm orb}=0.2$ day).
Although the nature of its CO (NS or BH) is still in debate,
its optical counterpart, V1521 Cyg, is a Wolf-Rayet star of the WN type \citep{Kerkwijk1992,Fermi2009}.
It means that V1521 Cyg is helium-rich star. Based on the ionisation
structure of the wind from Cygnus X-3, \cite{Terasawa1994} estimated that the mass for V1521 Cyg
was about $7^{+3}_{-2}M_\odot$. Compare to the typical mass of Wolf-Rayet stars, 
the progenitor of V1521 Cyg  must have lost enormous
mass via stellar wind or Roche lobe overflow (RLOF)\citep[e. g.,][]{Lommen2005}.
However, in our work, its progenitor may have undergone different evolutions.
As shown Figure \ref{fig:pfpi} in for the changes of pre- and post-CCSN orbital periods,
the progenitor system should have an orbital period shorter than $\sim 1$ day in order to form a HMXB
with a short orbital period similar to that of the Cygnus X-3 after the CCSN explosion.
We can estimate, based on Figure \ref{fig:umass}, that the stripped mass from the progenitor of V1521 Cyg could reach about 10 $M_\odot$
 during CCSN process. Therefore, the mass of Cyg progenitor might be $\sim 20 M_\odot$ implying that
  most of its hydrogen-rich envelope might be blown away
when the CO of Cygnus X-3 was formed.

Similarly, IWGA J0648-4119 also has a very short orbital period ($P_{\rm orb}=1.55$ days) implying that
its CO is likely a NS, but the massive white dwarf cannot be excluded\citep{Mereghetti2016}.
Its optical counterpart, HD 49798, is a hot subdwarf of O6 spectral
type with a mass of 1.50$M_\odot$\citep{Mereghetti2009}. Hot subdwarfs are core-helium-burning stars
with very thin hydrogen envelope whose mass is lower than 0.01 $M_\odot$ \citep{Heber2009}.
Hot subdwarfs in binary systems originate from the common-envelope ejection or stable RLOF\citep{Han2002}.
Based on Figure \ref{fig:pfpi}, our models predict that the pre-SN progenitor system should
have an orbital period shorter than $\sim 3$ day in order to form a HMXB with a
short orbital period similar to that of the IWGA J0648-4119 after the CCSN explosion.
It is possible that several $M_\odot$ was stripped away from the progenitor of HD 49798 when it evolved into the later phase in the MS
or the Hertzsprung gap. Considering the mass of HD 49798 is only about 1.50 $M_\odot$, we estimate that
its progenitor should have a mass of about $10 M_\odot$.

\section{Conclusions}
We have estimated the stripped masses from rotating stars based on the fitting
formula given by  \cite{Hirai2014} and \cite{Liu2015} together with the observational
data for HMXB orbital periods. Our results show that the amount of mass stripped is greatly dependent
on the orbital periods similar to that given in the previous literatures.
However, the rotational velocity introduces an uncertainty up to a
factor of about 2.   We focus on the evolutions of the rotational velocities,
and divide the evolutions into three phases: the impact, thermally adjusting  and normal phases.
We find that a Be star can evolve into a non-Be star if it is stripped with a mass higher than about 1 $M_\odot$.

Based on the observed orbital periods, we estimate that a mass of  several $M_\odot$
should have been stripped from V1521 Cyg and HD 49798.
They are the optical counterparts of Cygnus X-3 and  IWGA J0648-4119, respectively, and both are hydrogen depleted.
It is probable that the whole hydrogen-rich envelopes of their progenitors might have been stripped
when the COs form.

\section*{Acknowledgments}
GL thanks Dr Rai Yuen for polishing
the English language of the manuscript. This work was supported by
XinJiang Science Fund for Distinguished Young Scholars under No. 2014721015,
the National Natural Science Foundation
of China under Nos. 11473024, 11363005 and 11163005.

\bibliographystyle{aasjournal}
\bibliography{lglapj}


\end{document}